\documentclass{article}

\usepackage{PRIMEarxiv}

\usepackage[utf8]{inputenc} 
\usepackage[T1]{fontenc}    
\usepackage{hyperref}       
\usepackage{url}            
\usepackage{booktabs}       
\usepackage{amsfonts}       
\usepackage{nicefrac}       
\usepackage{microtype}      
\usepackage{lipsum}
\usepackage{fancyhdr}       
\usepackage{graphicx}       
\usepackage{amsmath}
\usepackage{amssymb}
\graphicspath{{media/}}     

\newcommand{\WI}[2]{#1_{\mathrm{#2}}}
\newcommand{\rISCO}{\WI{R}{ISCO}}
\newcommand{\rFill}{\WI{R}{fill}}
\newcommand{\rRoche}{\WI{R}{Roche}}
\newcommand{\mNS}{\WI{M}{NS}}
\newcommand{\mBH}{\WI{M}{BH}}
\newcommand{\rNS}{\WI{R}{NS}}

\title{Stripping of a Neutron Star in a Close Binary System in a Pair with a Black Hole
\thanks{\textit{\underline{Citation}}: 
Kramarev, N.I., Kuranov, A.G., Yudin, A.V., Postnov, K.A. Stripping of a Neutron Star in a Close Binary System in a Pair with a Black Hole. \textit{Astron. Lett.} \textbf{50}, 302–316 (2024). \textbf{DOI: 10.1134/S1063773724700166.}} 
}

\author{
  N.I.~Kramarev \\
  National Research Center ''Kurchatov Institute'', Moscow, 117218 Russia\\
  Sternberg Astronomical Institute, Moscow State University, Moscow, 119234 Russia \\
  \texttt{kramarev-nikita@mail.ru} \\
   \And
  A.G.~Kuranov \\
  Sternberg Astronomical Institute, Moscow State University, Moscow, 119234 Russia \\
   \And
   A.V.~Yudin \\
   National Research Center ''Kurchatov Institute'', Moscow, 117218 Russia\\
   Novosibirsk State University, Novosibirsk, Russia \\
   \And
   K.A.~Postnov \\
   Sternberg Astronomical Institute, Moscow State University, Moscow, 119234 Russia \\
}

\begin{document}
\maketitle

\begin{abstract}
We consider the final evolutionary stages of a neutron star–black hole pair. According to the current paradigm, such systems eventually coalesce, which in some cases is accompanied by neutron-star tidal disruption. Using analytical methods, we show that the scenario of slow (of the order of several seconds) neutron star stripping by the black hole is also possible, depending on the system parameters (the initial masses and intrinsic angular momenta of the components, the equation of state for the neutron
star). Reaching the lower mass limit (about one tenth of the solar mass), the neutron star explodes to produce a comparatively powerful electromagnetic transient. Our population calculations show that the stripping mechanism is possible in 50–90\% of the cases among all coalescing neutron star–black hole pairs, depending on the model assumptions about the evolution of close binary systems (the common-envelope
efficiency parameter, the supernova explosion mechanism) and the initial metallicity of the stellar population. Because of the large mass of the ejected material, the kilonova emission in this scenario has good prospects of detection.
\end{abstract}


\section{INTRODUCTION}

The mergers of binary neutron stars (NSs) and NSs with black holes (BHs) have attracted the attention of astrophysicists as hypothetical sources of powerful gravitational-wave (GW) and electromagnetic (EM) transients, in particular, gamma-ray bursts, for half a century \cite{LattimerSchramm1974,ClarkEardley1977,Blinnikov1984,Eichler1989}. The neutron-rich material ejected during such a merger is an appropriate place
for the synthesis of heavy elements. Therefore, it was hypothesized \cite{Li1998} that the nucleosynthesis process should be associated with the phenomenon of the so-called kilonova: a long-lived thermal transient radiating in the optical, infrared, and ultraviolet ranges of the EM spectrum \cite{Metzger2010}. These views received a reliable confirmation after August 17, 2017, when the GW event GW170817 and the gamma-ray burst GRB170817A from a pair of coalescing NSs were recorded almost simultaneously \cite{Abbott2017}. The emission from the kilonova AT2017gfo associated
with GW170817 was also soon detected \cite{Villar2017}.

Nevertheless, so far no source of EM emission from the mergers of NS–BH pairs has been found for sure \cite{Bhattacharya2019, Ekanger2023}. The estimates made by the population synthesis method do not give an optimistic prospect for the detection of kilonovae \cite{Drozda2022} and gamma-ray bursts \cite{Postnov2020} from such
mergers. According to the current paradigm of NS–BH mergers \cite{Kyutoku2021}, there are two reasons for this. First, the events of rapid tidal disruption of the NS before its plunge into the BH, whereby the ejection of material occurs, turn out to be rare compared to the events without tidal disruption. Second, the mass of the material ejected during NS disruption will be much lower than that during the merger of two NSs. For this reason, the mergers in a NS–BH pair give comparatively weak and difficult-to-detect EM transients.

Therefore, in addition to the widely discussed scenarios for the merger of NS–BH systems (accompanied or unaccompanied by rapid NS tidal disruption), here we also propose to consider the possibility of slow NS stripping by the massive component.

So far the stripping scenario has been developed only for NS–NS systems \cite{Blinnikov1984, Blinnikov2021, Blinnikov2022, KramarevYudin2023str}. Its essence
is as follows: as a NS approaches a more massive component (another NS or BH) because of the GW emission, the low-mass NS fills its Roche lobe and begins to spill over onto the more massive secondary component. Because of the accretion of material, the asymmetry of the system increases, and the components move away from each other. The stripping process proceeds on a comparatively long time scale (of the order of seconds) determined by the rate of angular momentum loss by the system carried away by GWs. As a result, the low-mass NS reaches the minimum NS mass limit (of the order of a tenth of the solar mass), loses the hydrodynamic stability, and explodes, producing a powerful EM transient and contributing to the cosmic synthesis of heavy elements
\cite{Blinnikov1990,Panov2020,Yudin2022,Yip2022}. Because of the fairly large mass of the ejected material, the NS stripping mechanism in a NS–BH binary system observationally turns out to be more promising than NS tidal disruption \cite{Kawaguchi2016}.\footnote{Some hydrodynamic simulations (see, e.g., \cite{Lovelace2013,Kiuchi2015,Hayashi2021,Hayashi2022}) demonstrate a significant mass of the ejected material. Nevertheless, these results were obtained for extremely large
specific intrinsic BH angular momenta, $\WI{\chi}{BH} \gtrsim 0.7$, and/or relatively small component mass ratios, $\mBH/\mNS \lesssim 3$, which agrees poorly with both present-day population calculations \cite{Xing2023} and GW observations of candidates for NS–BH systems \cite{Zhu2022}.} The main question — is this scenario possible in nature? And if so, how often is it realized compared to NS tidal disruption? This paper is devoted to the answers to these questions.

The plan of the paper is as follows: first we describe the conditions under which a particular scenario for the evolution of NS–BH systems is realized and analytically
determine the mass boundaries between the scenarios. Thereafter, we find the fraction of NS–BH systems that will be stripped at the end of their evolution by population synthesis methods. In the concluding sections we summarize our main results
and discuss the prospects for a further development of the problem.

\section{THE FINAL EVOLUTIONARY STAGES OF A NS–BH BINARY SYSTEM}

\subsection{The NS–BH Approach Stage}

\begin{figure}
	\centering
	\includegraphics[scale=0.7]{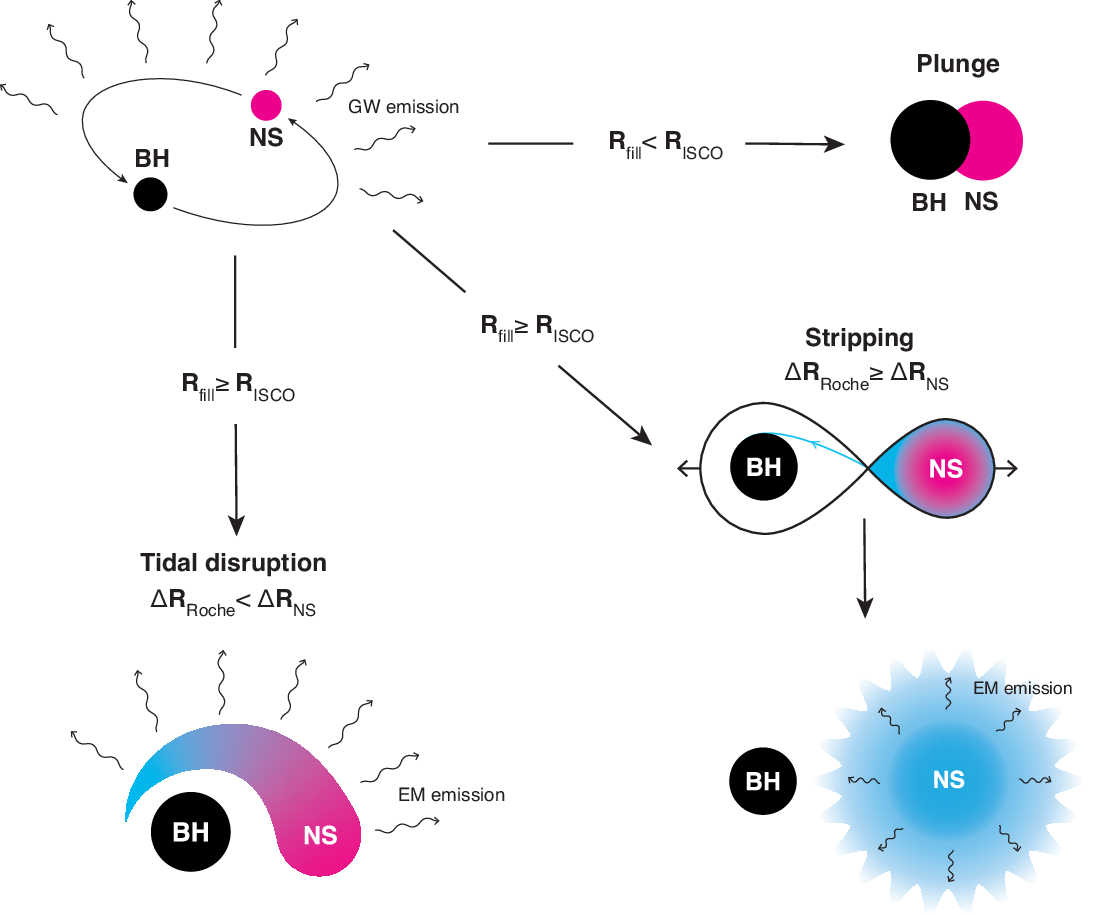}
	\caption{Three scenario for the final evolution of a NS–BH
		binary system: NS plunge, rapid tidal disruption of the
		NS before its plunge, and slow NS stripping (for details,
		see the text).}
	\label{Fig1}
\end{figure}

A NS and a BH with masses $\mNS$ and $\mBH$ in a close binary system approach each other for a long time due to the loss of energy and orbital angular
momentum through GW emission (see Fig.~\ref{Fig1}). Although the initial eccentricity of the system, $e$, can be fairly large \cite{Xing2023}, its value will be negligible by the onset of mass transfer under the influence of GW emission, $e\leq 10^{-3}$ \cite{Kowalska2011}. Therefore, here and below we will assume that the
components revolve in circular orbits, $e=0$.\footnote{An analytical study of the stripping process in a NS–BH system with a nonzero eccentricity can be found in \cite{Davies2005}.}

\subsection{The Ejection of Material During Mass Transfer}

The fundamental possibility of the ejection of material during the coalescence of a NS–BH system is determined by the ratio of two characteristic radii of the system. Let us first determine the radius of the innermost stable circular orbit around the BH, $\rISCO$, which depends on the specific intrinsic angular momentum (spin) of the BH, $\WI{\chi}{BH}{=}2\WI{J}{BH}/\WI{R}{g} \mBH c$, where $\WI{R}{g}{=}2G\mBH/c^2$ is the gravitational radius of the BH, and the inclination angle of the BH spin with respect to the orbital angular momentum of the system, $\WI{\theta}{BH}$. Generally, $\rISCO$ cannot be expressed analytically; its value can be found only numerically (see, e.g., \cite{Rezzolla2016}). Nevertheless, to terms $O(\WI{\chi}{BH}^2)$ the radius of the innermost stable circular orbit is equal to its value in a plane perpendicular to
the BH spin (at the equator):
\begin{equation}
	\rISCO(\WI{\chi}{BH},\WI{\theta}{BH}){=}\rISCO^{\mathrm{eq}}(l),
\end{equation}
where the BH spin projection, $l{=}\WI{\chi}{BH} \cos{\WI{\theta}{BH}}$, acts
as the sole argument. The radius of the innermost stable circular orbit at the equator is described by the following well-known analytical formula \cite{Bardeen1972}:
\begin{equation}
	\frac{\rISCO^{\mathrm{eq}}}{\WI{R}{g}} {=} \frac{3{+}Z_2{-}\text{sign}(l)\sqrt{(3{-}Z_1)(3{+}Z_1{+}2 Z_2)}}{2},  \label{r_isco}
\end{equation}
where we introduced auxiliary functions, $Z_1(l)=1{+}\sqrt[3]{1-l^2}\left[\sqrt[3]{1{+}l}+\sqrt[3]{1{-}l}\right]$ and $Z_2(l)=\sqrt{3 l^2{+}Z_1^2(l)}$. The dimensionless parameter $l$ and
the radius itself can change within the ranges $l \in [-1;+1]$ and $\rISCO^{\mathrm{eq}} \in [4.5 \WI{R}{g}; 0.5 \WI{R}{g}]$, respectively. Positive and negative values of $l$ correspond to the orbital motion in the direction of BH rotation and in the opposite direction, respectively. As the population synthesis \cite{Xing2023} and GW data processing \cite{Gompertz2022} results show, the values of
the angle $\WI{\theta}{BH}$ can differ greatly from zero, but, at the same time, the specific BH spin itself and its projection onto the orbital angular momentum axis
turn out to be close to zero.\footnote{Note that the so-called effective specific spin of the system is often used in the literature: $\WI{\chi}{eff}{=}(\mBH \WI{\chi}{BH} \cos{\WI{\theta}{BH}} {+}\mNS \WI{\chi}{NS} \cos{\WI{\theta}{NS}})/(\mBH{+}\mNS)$. Since the specific NS spin may be assumed to be zero with a great accuracy (see, e.g., Eq. (2) from \cite{Kyutoku2021}), $\WI{\chi}{NS}{=}0$, the effective specific spin will be defined by a more compact expression: $\WI{\chi}{eff} = l \mBH/(\mBH{+}\mNS)$.} Therefore, here and below for $\rISCO$ we will use Eq.~(\ref{r_isco}) with the argument $l{=}\WI{\chi}{BH} \cos{\WI{\theta}{BH}}$.

Finally, let us introduce the characteristic radius of the system at which the NS fills its Roche lobe:
\begin{equation}
	\rFill = a(\rNS{=}\rRoche), \label{r_fill}
\end{equation}
where $a$ is the distance between the components, and $\rNS(\mNS)$ is the NS radius dependent on its mass in accordance with the chosen equation of state (EoS). The Roche lobe radius, $\rRoche$, is parameterized in a standard way \cite{Eggleton1983}: 
\begin{equation}
	a = \rRoche / f(q,z), \label{r_Roche}
\end{equation}
where $q = \mNS /\mBH \leq 1$ is the component mass ratio that is a measure of asymmetry of the system. The dependence of the function $f$ on the parameter 
\begin{equation}
	z = \frac{2G\WI{M}{tot}}{c^2 a}, \label{z_rel}
\end{equation}
where $\WI{M}{tot}{=}\mNS{+}\mBH$ is the total mass of the system, takes into account the relativistic effects to the second order of the post-Newtonian (2PN) approximation \cite{Ratkovic2005}:
\begin{equation}
	f(q,z)=\frac{0.49 q^{2/3} \Big[1+z\big(1.951 q^{1/5}\!{-}1.812\big)\Big]}{0.6 q^{2/3}+\ln\big(1{+}q^{1/3}\big)}. \label{f_q}  
\end{equation}
In what follows, we will also need a more compact form of Eq.~(\ref{f_q}): $f(q,z){=}\WI{f}{Egg}(q)[1{+}z\WI{f}{z}(q)]$. Mathematically, the relativistic parameter $z$, which determines the contribution of the general relativity (GR)
effects, acts as a small parameter of the problem. Its physical meaning is also clear from Eq.~(\ref{z_rel}): the ratio of the gravitational radius of the system to the
distance between the components. Although the function $f(q,z)$ depends implicitly on the distance $a$, it is easy to show that $a$ can be explicitly expressed from Eqs.~(\ref{r_Roche})--(\ref{f_q}). 

Naturally, the ejection of material in the NS–BH system turns out to be possible only if the NS will fill its Roche lobe without first plunging into the BH, i.e., $\rFill \geq \rISCO$ (see Fig.~\ref{Fig1}). In accordance with the aforesaid, this is determined by the initial masses of the components, the BH spin projection, and the NS EoS.

\subsection{The Stability of Mass Transfer and NS Stripping}
Let us turn to the central theme of our work, namely the possibility of long NS stripping. Let the NS fill its Roche lobe as a result of the approach of the components and undergo no plunge into the BH. For the onset of stable mass transfer it is necessary
that the size of the NS Roche lobe grows faster than its radius during mass transfer:\footnote{Recall that in the range of low masses the NS is an object with a negative exponent of the mass–radius relation.} 
\begin{equation}
	\WI{\dot{R}}{Roche} \geq \WI{\dot{R}}{NS}. \label{stab0}  
\end{equation}
In the case of violation of this condition, the NS will be tidally disrupted with the subsequent formation of a disk-shaped or crescent-shaped structure around the BH (see Fig.~\ref{Fig1}). Using the equation for the change in the orbital angular momentum of the system that goes into the GW emission, the following stability condition can be obtained in the simplest case: $q\leq 2/3$ \cite{Jaranowsk1992J, PortegiesZwart1998}.\footnote{This stability criterion was first obtained by B.~Paczynski with co-authors \cite{Paczynski1969} for stars with deep convective envelopes.} Allowance for the change in the intrinsic angular momenta of the components can give an even more optimistic result (see \cite{KramarevYudin2023str} for NS–NS systems). Note also that the analytical estimate obtained in Appendix~C.1 of the review \cite{Kyutoku2021} dismisses any possibility of stable mass transfer in NS–BH systems. This is because the authors erroneously considered the case of extremely compact orbits and used a simpler formula for the distance between the components: $\rFill \lesssim \rNS \, q^{-1/3}$. Moreover, no GR effects were taken into account in any of the mentioned analytical estimates.

In this paper we will take into account the relativistic effects in the 2PN approximation, since the relativistic parameter $z$ changes at a level of several tens of percent. Let us substitute the parametrization (\ref{r_Roche}) into the stability criterion (\ref{stab0}). Given the condition for the mass transfer to be conservative, $\WI{M}{tot}{=}const$, it transforms to the following form:
\begin{equation}
	\frac{\dot{a}}{a}\left[1{-}\frac{d \ln f}{d \ln z}\right]\geq \frac{\WI{\dot{M}}{NS}}{\mNS}\left[\frac{d \ln \rNS}{d \ln \mNS}{-}\frac{d \ln f}{d \ln q}(1{+}q)\right]. \label{stab1}  
\end{equation}
To relate the change in the distance $\dot{a}$ to the corresponding change in the donor mass $\WI{\dot{M}}{NS}$, we will use the equation for the change in the total angular
momentum of the system:
\begin{equation}
	\WI{\dot{J}}{GW}=\WI{\dot{J}}{orb}+\WI{\dot{J}}{BH}, \label{J_tot}  
\end{equation}
where we neglected the intrinsic angular momentum of the NS in view of its smallness. Equation~(\ref{J_tot}) also defines the long-term evolution of the system in the stripping mechanism. This stripping proceeds on the time scale determined by the loss of angular momentum of the system through GW emission (see, e.g., \cite{Paczynski1967}):
\begin{equation}
	\WI{\dot{J}}{GW}=-\frac{\sqrt{32}}{5} \nu \WI{J}{orb}^{\mathrm{N}} \WI{\Omega}{orb}^{\mathrm{N}} z^{2.5}, \label{J_GW}  
\end{equation}
where $\WI{J}{orb}^{\mathrm{N}}$ and $\WI{\Omega}{orb}^{\mathrm{N}}$ are the orbital angular momentum of the system and the orbital frequency of the components in the Newtonian approximation. Here we also introduced the dimensionless combination of the mass ratio $\nu{=}\mBH\mNS/\WI{M}{tot}^2$ that changes within the range $\nu \in [0;0.25]$. It can be clearly seen from the form being used by us that the GW emission is a 2.5PN-order effect, and, therefore, Eq.~(\ref{J_GW}) requires no refinement. 

For our further analysis of the binary system parameters needed to us to 2PN terms we will use the results from \cite{SchaferWex1993,Wex1995}. For example, the orbital angular momentum of the system can be written as follows:
\begin{equation}
	\WI{J}{orb}{=}\WI{J}{orb}^{\mathrm{N}}\left[1{+}z{-}\frac{3}{2\sqrt{8}}\left(2{+}\frac{3}{2}q\right)\frac{\nu l}{q} z^{1.5}{+}\frac{42{-}43\nu}{64}z^2\right], \label{J_orb}  
\end{equation}
where the term at $z^{1.5}$ takes into account the coupling between the intrinsic angular momentum of the BH and the orbital angular momentum of the system, the so-called \textit{spin–orbit coupling}.

Finally, let us turn to the analysis of the term $\WI{\dot{J}}{BH}$ that takes into account the accretion spin-up of the massive component that was studied in detail by us previously in the Newtonian approximation \cite{KramarevYudin2023acc}. The term $\WI{\dot{J}}{BH}$ can be written as follows:
\begin{equation}
	\WI{\dot{J}}{BH}=-\WI{\dot{M}}{NS} \mathfrak{j} a^2 \WI{\Omega}{orb}^{\mathrm{N}}. \label{J_BH}  
\end{equation}
The function $\mathfrak{j}$, i.e., the specific angular momentum of the accreting matter, depends on the ratio of the NS mass to the total mass of the system, $\mNS/\WI{M}{tot}$, and the dimensionless stopping radius, $\WI{r}{st}$. In general,
two accretion regime can take place during the stripping of the low-mass component \cite{LubowShu1975}. In one case, the stream of accreting matter hits the accretor surface. If, however, the minimal approach of the stream of accreting matter $\WI{R}{m}$ is greater than the equatorial radius of the accretor, then an accretion disk with an outer radius $\WI{R}{d}$ is formed. For the described accretion regimes the stopping radius is
\begin{equation}
	\WI{r}{st} = 
	\begin{cases}
		\WI{R}{st}/a, \; \WI{R}{st}\geqslant R_{\mathrm{m}}, \\
		R_{\mathrm{d}}/a, \; \WI{R}{st}<R_{\mathrm{m}}.
	\end{cases} \label{r_stop}
\end{equation}
The approximations for the functions $\mathfrak{j}$, $\WI{R}{m}$ and $\WI{R}{d}$ were obtained in \cite{KramarevYudin2023acc} in the Newtonian approximation. We will use these approximations in GR to calculate the BH spin-up following the approach from \cite{KramarevYudin2023str}. For this purpose, we will introduce the so-called \textit{effective BH stopping radius}, $\WI{R}{st}$, at which the accreting matter
will transfer its angular momentum to the BH. In fact, the BH stopping radius being introduced by us is the sole artificial parameter of the problem. Using it at this stage of our studies allows us not to solve the restricted three-body problem in GR, which, however, should be done in subsequent papers.

From Eqs.~(\ref{stab1})--(\ref{J_BH}) we can find the transfer stability criterion in the 2PN approximation. However, since it is cumbersome, here we will present this criterion only in the 1PN approximation, where the relativistic corrections are linear in $z$:
\begin{eqnarray}
	\frac{d \ln \rNS}{d \ln \mNS} \geq (1{+}q)\left[\frac{d \ln \WI{f}{Egg}}{d \ln q}+zq\frac{d \WI{f}{z}}{d q}\right] -2(1{-}q)\left[1{+}z(2{-}\WI{f}{z})\right]+2\mathfrak{j} (1{+}q)\left[1{+}z\left(1{-}\WI{f}{z}\right)\right].
	\label{stab_PN1}  
\end{eqnarray}
The corresponding expression in the 2PN approximation is given in the Appendix. Note that in comparison with Eq.~(\ref{stab_PN1}), the stability criterion obtained in
the 2PN approximation (see the Appendix, Eq.~(\ref{Appendix_stab_PN2})) also contains the dependence on the projection of the specific BH spin onto the orbital angular momentum axis, $l$, and the corresponding cosine of the inclination angle, $\cos{\WI{\theta}{BH}}$. This is related to the above-mentioned spin–orbit coupling effect.

Thus, if the NS–BH system with some initial parameters satisfies the derived inequality, then the NS, having filled its Roche lobe, will be slowly stripped by the BH. At some time the transfer stability is lost, the NS will reach its minimum mass limit and will explode \cite{Blinnikov2021}.

\section{THE MASS BOUNDARIES BETWEEN VARIOUS SCENARIOS}

\subsection{The Influence of Relativistic Effects}

\begin{figure}[h]
	\centering
	\begin{minipage}[h]{0.49\linewidth}
	\includegraphics[width=\columnwidth]{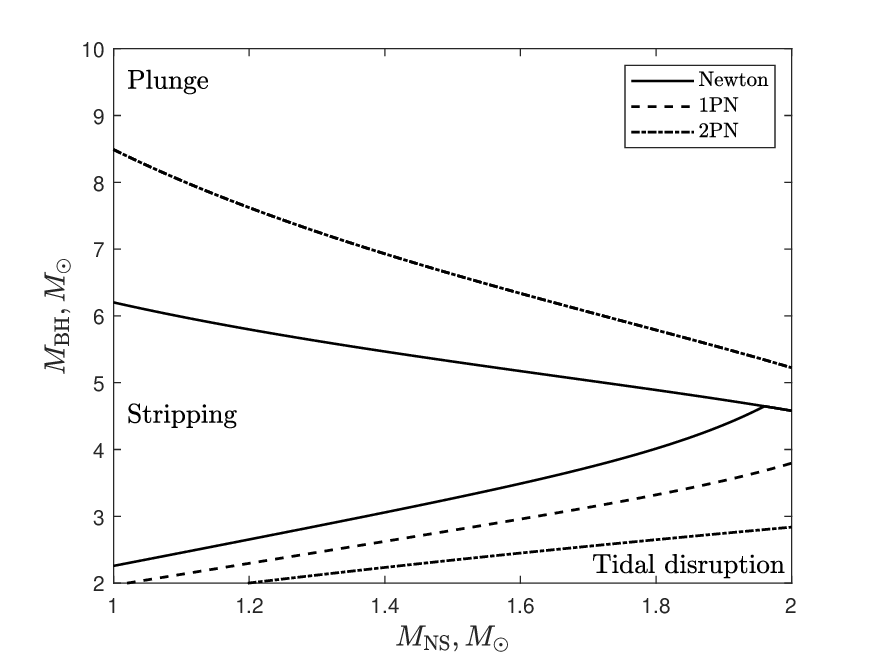}
	\caption{The boundaries between the scenarios for the final
		evolution of the NS–BH system for various component
		masses in the purely Newtonian, first, and second post-
		Newtonian approximations. By default, all our calculations
		were carried out for the following parameters: $l=0$,  $\WI{R}{st}=\WI{R}{g}$, and EoS BSk22 (for details, see the text).}
	\label{Fig2}
\end{minipage}
\hfill
\begin{minipage}[h]{0.49\linewidth}
	\includegraphics[width=\columnwidth]{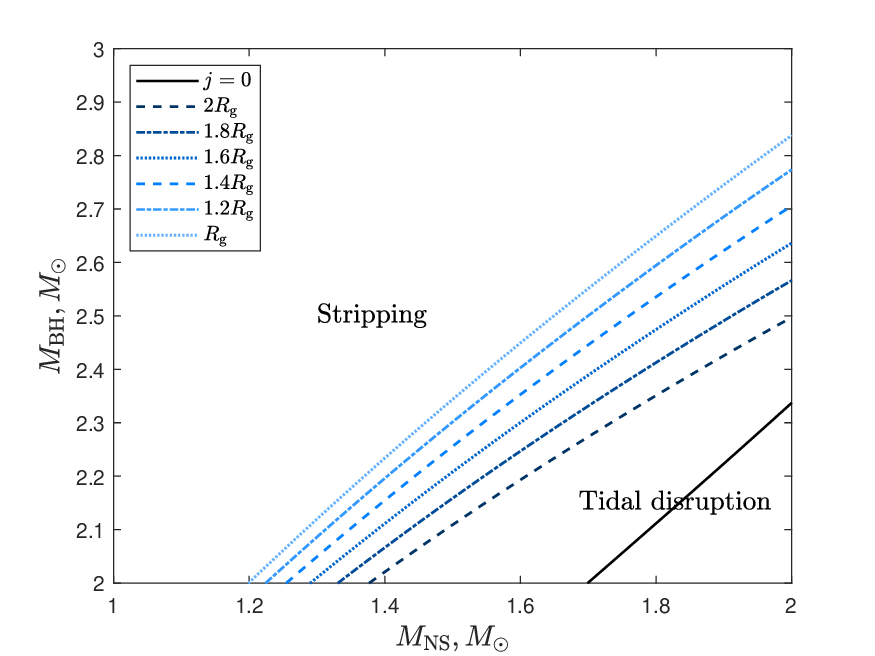}
	\caption{The mass boundaries between the NS stripping
		and tidal disruption scenarios for various values of the
		effective BH stopping radius, $\WI{R}{st}$. The black solid line
		corresponds to the calculation without the BH accretion
		spin-up effect. All our calculations were carried out in the
		2PN approximation with $l=0$ and EoS BSk22.}
	\label{Fig3}
\end{minipage}
\end{figure}

Given all of what has been said above, we can determine the domains of initial parameters of the components at which a particular scenario for the final evolution of the NS–BH system is realized and can estimate the influence of various effects. Let us
first discuss the major innovation of our work compared to the previous analytical calculations, namely the contribution of GR effects. Figure~\ref{Fig2} presents the mass boundaries between the NS merging, tidal disruption, and stripping scenarios in the Newtonian (solid line) and second post-Newtonian (dash–dotted line) approximations. For definiteness, we everywhere set the initial BH spin equal to zero — we will discuss its influence on the positions of the boundaries below separately. Curiously, the inclusion of relativistic effects increases the domain of initial NS and BH masses at which the stripping scenario is realized. This is because the orbital angular momentum of the system defined by Eq.~(\ref{J_orb}) in the 1PN and 2PN orders is greater than the purely Newtonian expression. It turns out that by the time the NS fills its Roche
lobe, other things being equal, the total angular momentum of the system will be greater, increasing the mass transfer stability in such a system. At the same time, applying the relativistic correction for the effective Roche lobe size (\ref{f_q}) leads to an increase in the distance between the components, $\rFill$, at which the NS fills its Roche lobe, increasing its chances to avoid the plunge into the BH. As a result, all of this expands the domain of initial component masses at which the stripping scenario is realized. The boundary between the stripping and tidal disruption scenarios in the first post-Newtonian approximation (dashed line) according to the analytical criterion (\ref{stab_PN1}) is also drawn in Fig.~\ref{Fig2} for comparison.\footnote{Note that the boundaries between the NS plunge and stripping scenarios in Fig.~\ref{Fig2} in the 1PN and 2PN approximations coincide, since the relativistic formula for the Roche lobe size (\ref{f_q}) from \cite{Ratkovic2005} being used by us is simultaneously linear in $z$ and has the 2PN order.} As it must be, this boundary lies between the 2PN and purely Newtonian approximations. Note here that the position of the lower boundary for the stripping scenario in the Newtonian limit for $\mNS=1.4M_{\odot}$ in Fig.~\ref{Fig2} is consistent with the results of \cite{Davies2005}.

We also investigated the influence of the BH accretion spin-up on the position of the mass boundary between the NS tidal disruption and stripping scenarios. We took into account the contribution of GR effects in the purely Newtonian formula~(\ref{J_BH}) by
substituting various stopping radii, $\WI{R}{st}$. In this paper we are primarily interested in the influence of the effects being studied on the fraction of NS stripping
events. As can be seen from Fig.~\ref{Fig3}, the domain of initial masses at which the stripping mechanism is realized also decreases with decreasing $\WI{R}{st}$. This is
because for relatively small component mass ratios the fraction of the specific angular momentum of the accreting matter being transferred increases with decreasing stopping radius (see the solid blue and green lines in Fig.~9 from \cite{KramarevYudin2023acc}). Hence, the bulk of the system’s orbital angular momentum will be lost in a unit time, which, as a result, leads to a decrease in the mass transfer stability. To estimate a lower limit for the fraction of stripping events, below we will assume, by default, the BH stopping radius to be $\WI{R}{st}=\WI{R}{g}$.

\subsection{The Contribution of the Intrinsic BH Angular Momentum}

\begin{figure}
	\centering
	\begin{minipage}[h]{0.49\linewidth}
	\includegraphics[width=\columnwidth]{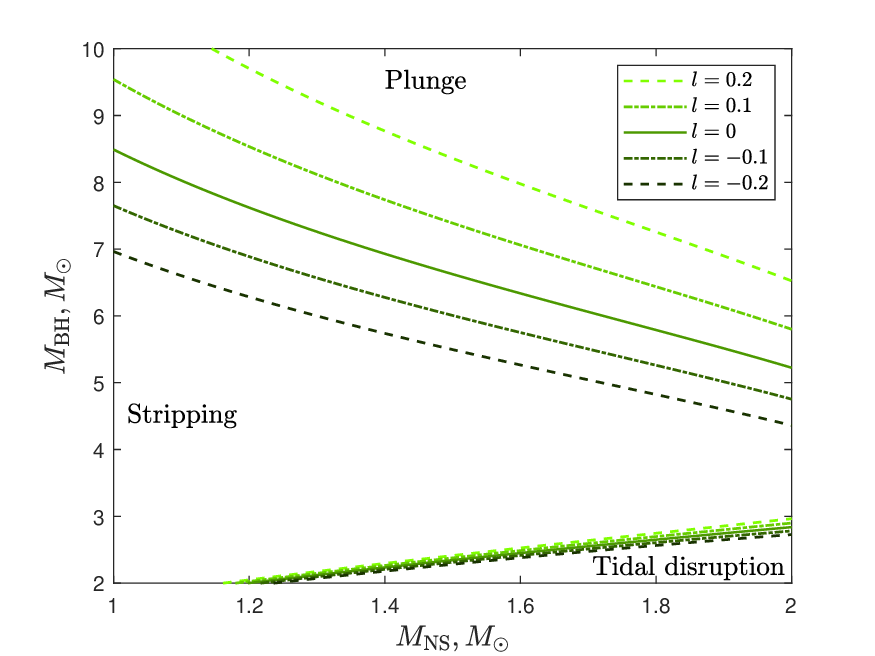} %
		\caption{The mass boundaries between the scenarios for
			various values of the projection of the specific BH angular
			momentum, $l$. The intrinsic BH angular momentum, by
			default, is everywhere assumed to be parallel to the orbital
			one, i.e., $\cos{\WI{\theta}{BH}}=1$. The remaining parameters are the
			same as those for Fig.~\ref{Fig1} and \ref{Fig2}.}
	\label{Fig4}
	\end{minipage}
	\hfill
	\begin{minipage}[h]{0.49\linewidth}
	\includegraphics[width=\columnwidth]{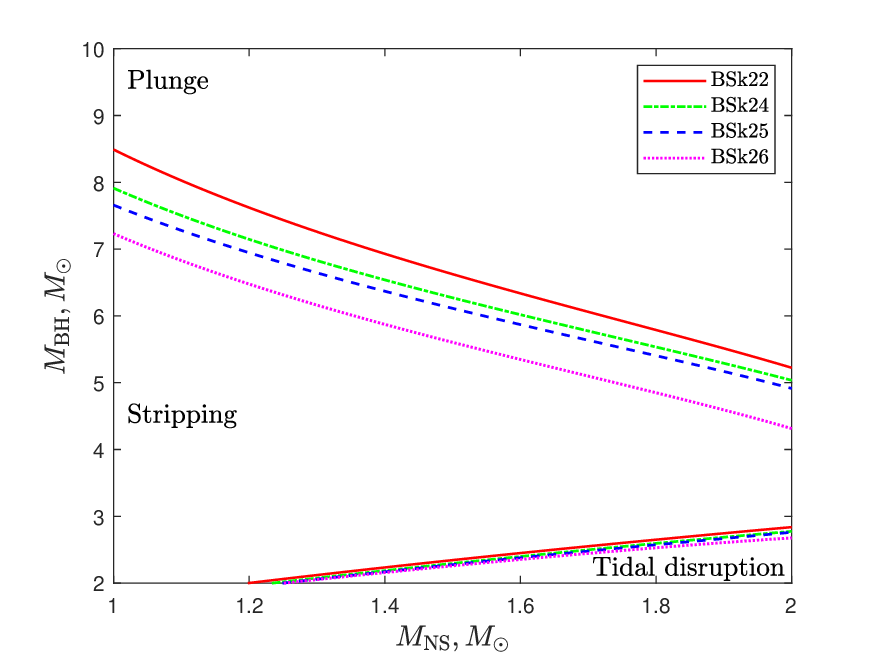} %
		\caption{The mass boundaries between the scenarios for various NS EoS (for details, see the text).}
	\label{Fig5}
	\end{minipage}
\end{figure}

Let us now briefly discuss the influence of the spin–orbit coupling dependent on the projection of the specific BH spin that we set above equal to zero. As follows from Fig.~\ref{Fig4}, the domain of initial component masses at which the NS completely plunges into the BH decreases with increasing $l$, consistent with hydrodynamic simulations \cite{Kyutoku2015} and similar analytical estimates (see Eq. (9) from \cite{Kyutoku2021}). At the same time, the spin–orbit correction affects weakly the mass transfer stability: the position of the corresponding mass boundary hardly changes for various $l$.

\subsection{The Role of the NS EoS}

We separately investigated the dependence of the positions of the mass boundaries on the NS EoS, which determines the NS radius appearing in Eqs.~(\ref{r_fill}) and (\ref{r_Roche}) and its logarithmic derivative with respect to the mass that enters into the stability criterion (Eqs.~(\ref{stab_PN1}) and (\ref{Appendix_stab_PN2})). As our calculations showed (see Fig.~\ref{Fig5}), the position of the boundary between the
NS stripping and tidal disruption scenarios is insensitive to the specific form of EoS taken from \cite{Pearson2018}. This is because in the domain of moderate masses the NS radii for various EoSs change within $\pm 1$ km, and the corresponding logarithmic
derivative is approximately equal to zero. Moreover, so small variations of the radius are a peculiarity of all the present-day EoSs that satisfy astrophysical observations and nuclear data \cite{Raaijmakers2020,Greif2020}. Note that previously we came
to an analogous conclusion when determining the mass boundary for NS–NS systems \cite{KramarevYudin2023str}. In this case, the position of the mass boundary associated with the NS plunge scenario is more sensitive to the EoS. Here our results are consistent with the semi-analytical calculations from \cite{Taniguchi2008}.

As an intermediate result, it can be said with confidence that within the approach being developed the scenario of slow stripping of a NS in a pair with a BH is possible in nature. Moreover, it is realized for a fairly extensive domain of initial parameters of the NS–BH system compared to the NS tidal disruption scenario. Therefore, its contribution to the total fraction of potentially detectable EM transients from NS–BH pairs can be dominant.

\section{POPULATION CALCULATIONS OF THE FRACTION OF THE STRIPPING MECHANISM}

For quantitative predictions of the fraction of the stripping mechanism we need to additionally know the distributions of initial parameters of NS–BH systems
determined by the evolution of close binary systems \cite{Postnov2014}. Recall that
the main parameters of interest to us are the initial masses of the components, $\mNS$ and $\mBH$, the specific intrinsic BH angular momentum, $\WI{\chi}{BH}$, and its
cosine of the inclination angle with respect to the orbital angular momentum of the system, $\cos{\WI{\theta}{BH}}$. To find them, we will use the \texttt{BSE} (Binary Stellar Evolution) population code \cite{Hurley2000,Hurley2002} modified to calculate the rotations of stellar cores by taking into account the main processes of the evolution of close binary systems \cite{Postnov2019,Postnov2020}. Let us list the main sources of uncertainties in the evolution of binary systems of massive stars that arise in all such calculations (see also \cite{Broekgaarden2021,Zhu2022,Xing2023}). These primarily include the so-called common-envelope efficiency parameter, $\WI{\alpha}{CE}$, equal to the ratio of the binding energy of the stellar core
and envelope after the main sequence (MS) to the orbital energy of the binary system before the onset of the common-envelope stage \cite{Webbink1984,Iben1984}. Another important source of uncertainties is the model of the distribution of NS and
BH masses after the supernova (SN) explosion (see, e.g., \cite{Fryer2012}). The stellar evolution result is also sensitive to the initial metallicity,  $Z$, which determines the stellar wind intensity \cite{Kudritzki2000} and the rate of stellar evolution. All of this determines both the specific form of the distribution of NS–BH systems by masses and spins of interest to us and the total number of NS–BH pairs with respect to the systems of other types: NS–NS, BH–BH, etc. The influence of each mentioned source of uncertainties is inseparably linked with the corresponding evolutionary stages of the system: the mass transfer between the components and the formation of compact objects. Their discussion is presented below.

\subsection{The Primary Mass Transfer Stage}

\begin{figure*}[h]		
	\centering
	\includegraphics[width=\textwidth]{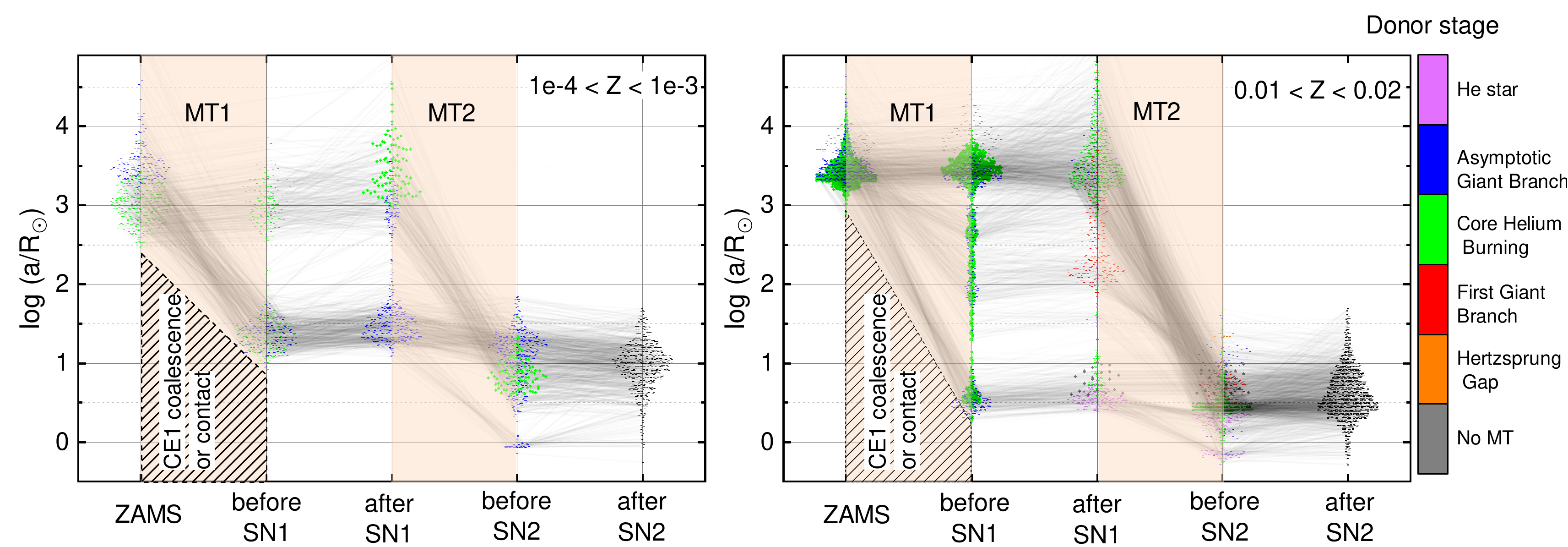}
	\caption{Evolution of the orbital semimajor axes of binary systems leading to the formation of NS–BH systems. The distributions on the zero-age main sequence (ZAMS) and immediately before and after the supernova explosions (before SN, after SN) are shown. The color scale reflects the evolutionary stage at which the donor at the Roche lobe filling time is. The hatching highlights the region at the primary mass transfer stage (MT1) in which the binary components coalesce either as a result of the onset of the common-envelope stage (CE1 coalescence) or because of the contact of the stars (contact). Left-hand panel: low-metallicity ($10^{-4}<Z<10^{-3}$) stars. The primary and secondary mass transfer stages (MT1 and MT2, respectively) occur exclusively in the donor core or shell helium burning episodes. Right-hand panel: stars of nearly solar chemical composition ($0.01<Z<0.02$). In the secondary mass transfer episodes the donor can be in virtually any evolutionary status between the post-MS star and the helium star (for details, see the text).}
	\label{Fig6}			
\end{figure*}

During its evolution the binary experiences several episodes of the interaction between the components that lead to the mass transfer stages (MT1 and MT2, see Fig.~\ref{Fig6}). The mass transfer stability, other things being equal (the component mass ratio, the degree of conservatism, etc.), depends on the evolutionary stage of the star by the Roche lobe filling time. This is determined mainly by two factors: the rate of stellar evolution (depends on $Z$) and the system geometry (the Roche lobe size).

In very massive stars (in our case, these are BH progenitors) the primary component fills its Roche lobe at the helium burning stage (core helium ignition begins almost immediately after hydrogen burnout) virtually independently of $Z$.

If the binary is not too wide, the donor manages to lose the bulk of its hydrogen–helium envelope (to the point of the exposure of the helium cores) at the
stage of stable mass transfer (SMT1) — the so-called case B according to the classification from \cite{Kippenhahn1967}. As a result, a system of a helium star (a WR star that subsequently collapses into a BH) and a secondary MS component that accumulated an additional mass at the accretion stage is formed.\footnote{Note that the accretor star responds to an increase in its mass by expansion. For initially close binary systems this can lead to contact of the stars at the mass transfer stage with the formation of a common envelope and the subsequent coalescence of the binary components.}

In initially wider binary systems the Roche lobe is filled by the primary component at later stages — a convective envelope has time to be formed in the star. In this case (the so-called case C) the mass transfer stage ends with the common-envelope phase (CE1). The remnants of such systems, unless they coalesce at the CE1 stage (the corresponding domain of parameters is indicated in Fig.~\ref{Fig6} by the hatching),
are close binaries with a helium star and a MS companion with a mass close to the initial mass on the ZAMS.

To describe the common-envelope stages, we used the $\alpha{-}\lambda$ formalism \cite{Webbink1984,Iben1984,deKool1990}. When calculating the envelope binding energy, $\Delta E_\mathrm{env}{=}GM_\mathrm{env}M_\mathrm{core}/(\lambda R)$, $\Delta E_\mathrm{env}$ is directly calculated using the open code of \cite{Loveridge2011}.

As the common-envelope efficiency increases (corresponding to lower values of the parameter $\WI{\alpha}{CE}$), the fraction of survived binary systems that passed
through the CE stage decreases (increasingly wider binary systems are subjected to coalescence). Thus, the common-envelope efficiency affects primarily the overall formation rate of coalescing NS–BH systems. In this case, the formation rate also depends on a set of additional scenario assumptions and can lie in a wide range (see \cite{Broekgaarden2021} and references therein): $(10^{-7}{-}10^{-5})\times \mathrm{SFR}\, [\text{yr}^{-1}]$, where SFR is the star formation rate in the Galaxy. At the same time, the influence of the parameter $\WI{\alpha}{CE}$ on the general form of the distributions in component masses and effective system spin and, eventually, on the fraction of the stripping mechanism must be not so significant.

\begin{figure*}[ht]
	\centering
	\includegraphics[width=\textwidth]{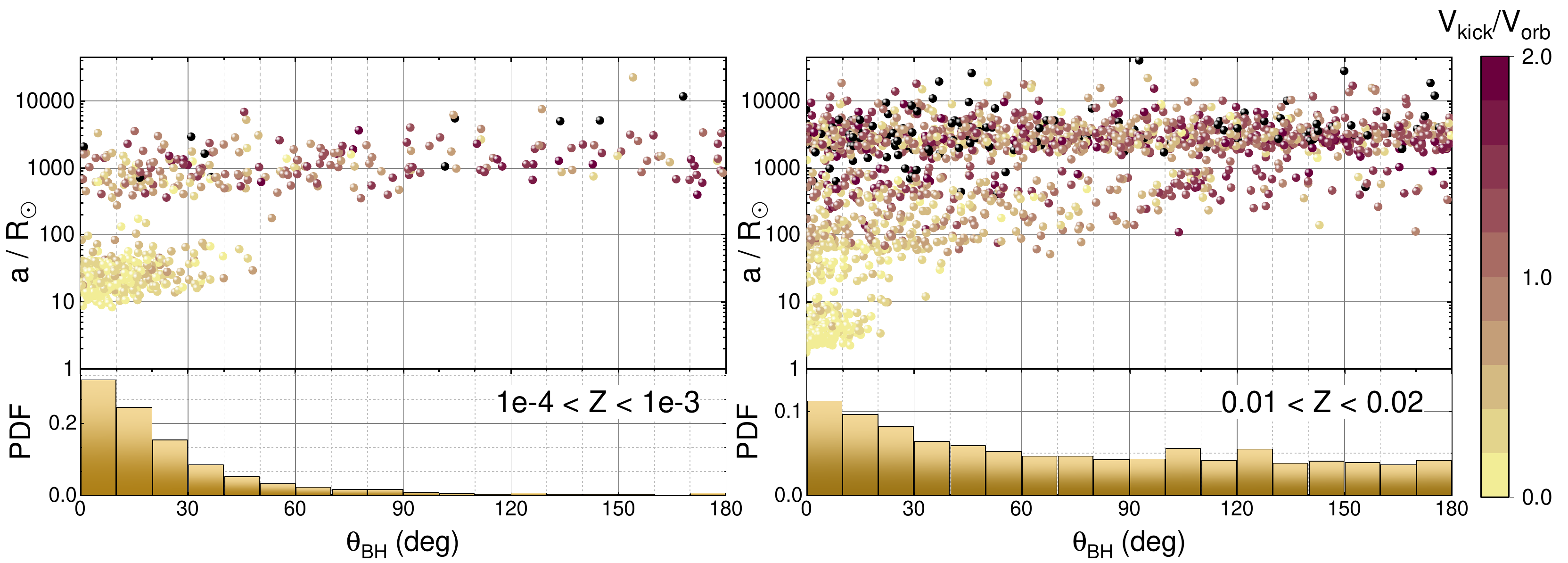}
	\caption{
		Top panels: the distribution of binary systems by orbital semimajor axes before the BH formation and by BH spin inclination angles ($\WI{\theta}{BH}$). The density of points is proportional to the number of systems. The color represents the ratio of the kick velocity and the orbital velocity of the pre-supernova. Bottom panels: the histograms illustrating the distribution of systems by BH spin inclination angles.  
		Left panels: low-metallicity ($10^{-4}<Z<10^{-3}$) stars. The fraction of close binaries ($10<a<100$) for which the additional kick velocity is low compared to the orbital one and has no significant influence, the
		rotation of the orbital plane is insignificant, is great. Right panels: stars of nearly solar chemical composition ($0.01<Z<0.02$), where
		the fraction of wide systems is fairly large. In this case, $\WI{\theta}{BH}$ is determined mainly by the additional kick velocity randomly directed in space: the distribution by rotation angles is nearly uniform.}
	\label{Fig7}
\end{figure*}

\subsection{The Secondary Mass Transfer Stage}

In contrast to the primary component that always fills its Roche lobe at the helium burning stages, the evolutionary status of the secondary one (the less massive component — the NS progenitor) in the secondary mass transfer episode directly depends on $Z$. For low-metallicity stars the secondary component also fills its Roche lobe at the helium burning stages. As $Z$ grows, the fraction of stars filling the Roche lobe
at increasingly early evolutionary stages increases. For nearly solar chemical composition the fraction of donors at the Hertzsprung gap and shell helium burning (First Giant Branch in Fig.~\ref{Fig6}) stages turns out to be significant. In addition, the evolutionary channel with repeated Roche lobe filling of the already evolved helium (He) star also becomes possible.

Such a difference in the donor types at the MT2 stages also leads to a significant difference in the evolutionary scenarios that give rise to NS–BH systems for different $Z$. Without going into details, note that a consequence of this, among other things, is also a change in the form of the distribution of systems by semimajor axes before the first supernova explosion (before SN1 in Fig.~\ref{Fig6}). At a low metallicity there is a distinct separation into ''close'' pre-supernovae (the dominant channel, $10R_\odot < a <100 R_\odot$) from the initially wide binaries that survived after the CE1 stage and ''wide'' pre-supernovae ($100 R_\odot < a <10^4 R_\odot$) from the systems that passed the stage of stable mass transfer (SMT1). In binaries with nearly solar chemical composition the form of the distribution is more complex and lies in a wider range.

At the same time, the donor type at the MT2 stage has virtually no effect on its outcome. Because of the large mass ratio (a BH with a massive star — the NS progenitor), the stage ends with a high probability with the formation of a common envelope (CE2) and a significant approach of the components (see Fig.~\ref{Fig6}).

\subsection{The Formation of Compact Objects}

\begin{figure*}[h]
	\centering
	\includegraphics[width=\textwidth]{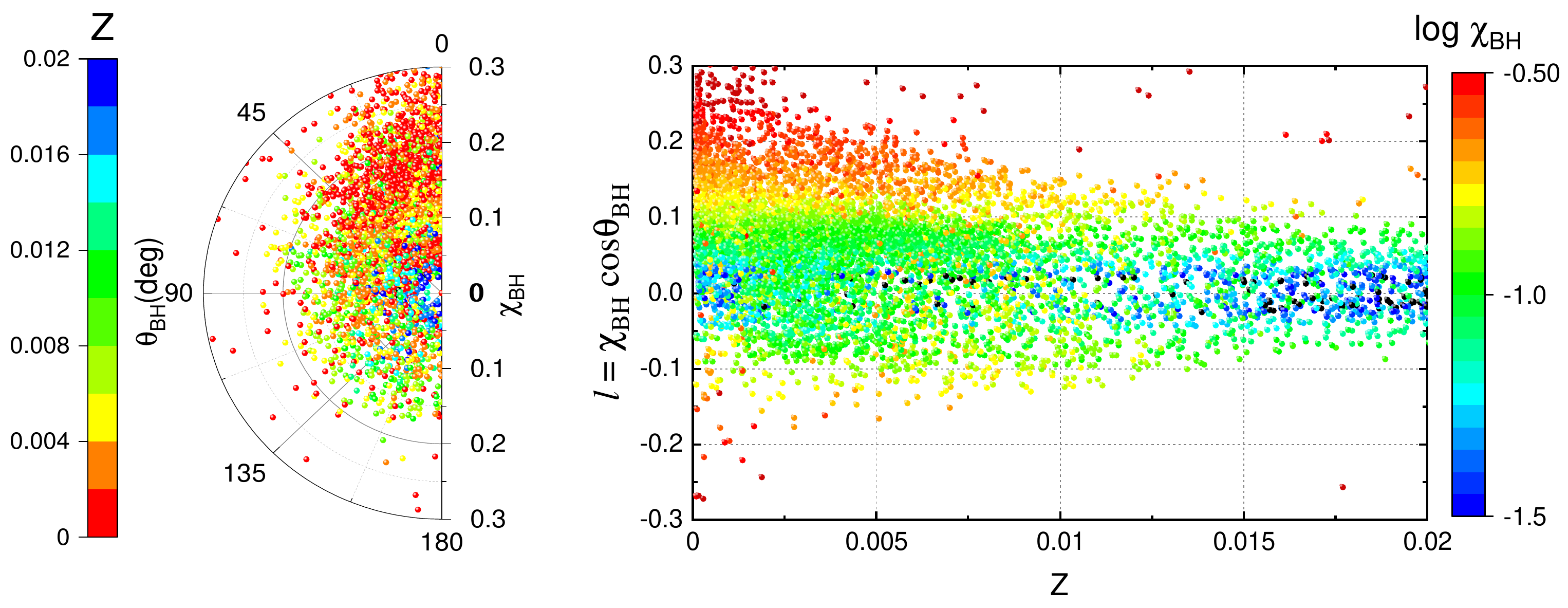}
	\caption{Left-hand panel: the distributions by intrinsic specific angular momenta ($\WI{\chi}{BH}$) and BH spin inclination angles to the orbital angular momentum of the system (${\WI{\theta}{BH}}$). The color scale reflects $Z$. Right-hand panel: the distribution of specific BH spins ($l = \WI{\chi}{BH}\cos{\WI{\theta}{BH}}$) by $Z$. The color scale reflects $\WI{\chi}{BH}$.}
	\label{Fig8}
\end{figure*}

The formation of a compact object (a BH or a NS) is accompanied by the acquirement of an additional velocity (the so-called kick). In our calculations we assume that the direction of the kick is isotropic in space and the magnitude of the velocity acquired by the BH during the collapse of a star with a mass $M_\mathrm{fin}$ is 
\begin{equation}
	V_{\rm kick} = \frac{M_\mathrm{fin}-\mBH}{M_\mathrm{fin}-M_\mathrm{Fe}}\,A_{\rm kick}, \label{e:vkick}
\end{equation}
where $M_\mathrm{Fe}$ and $M_\mathrm{BH}$ are the masses of the pre-supernova iron core and the BH, respectively, and the amplitude $A_{\rm kick}$ obeys the Maxwellian velocity
distribution with a dispersion $\sigma=265 \, \text{km} \, \text{s}^{-1}$ proposed for radio pulsars \cite{Hobbs2005}. For the NS the magnitude of the kick velocity does not
depend on the masses, i.e., $V_{\rm kick} = A_{\rm kick}$.\footnote{In a narrow range of initial masses close to the lower mass limit for the NS formation after the first mass transfer the evolution of the helium core presumably ends with an electron-capture SN (ECSN, see \cite{Miyaji1980,Siess2018}), which is accompanied by a small magnitude of the kick velocity and lower masses of the forming NSs (in our calculations we took $30 \, \text{km} \, \text{s}^{-1}$ and $1.25 \, M_\odot$, respectively). In this case, we found that neither the range of masses for ECSN nor the low kick velocity affect noticeably the results of our calculations.} The ratio of the kick velocity and the orbital velocity of the pre-supernova determines the measure of rotation of the orbital plane relative to the intrinsic angular momentum of the compact object and, accordingly, the angle $\WI{\theta}{BH}$.\footnote{Note that the second supernova explosion does not have a so serious effect on the rotation of the orbital plane, since after the secondary mass transfer stage all systems are close ones (the progenitors of NS–BH pairs).} As noted above, depending on the pattern of mass transfer at the primary mass transfer stage, two evolutionary scenarios can be identified:
\begin{itemize}
	\item
	in the CE1 scenario (the progenitors of close binary systems before the supernova explosion) the magnitude of the kick velocity is, as a rule, low with respect to the orbital one and, therefore, the rotation of the orbital plane is small. The distribution by angles has a distinct maximum near zero and monotonically falls off
	at larger angles. Such a form of the distribution is typical for low-metallicity stars (see the left-hand panels of Fig.~\ref{Fig7}); 
	\item
	in the SMT1 scenario (wide binary systems) the magnitude of the kick velocity is comparable to or exceeds the orbital one. The distribution in orbital plane rotation angles is almost flat. Such a form of the distribution is typical
	for stars of nearly solar chemical composition (see the right-hand panels of Fig.~\ref{Fig7}).
\end{itemize}

The chemical composition of the stars also affects the form of the distribution by the system’s specific effective spin (Fig.~\ref{Fig8}). A lower mass loss rate due
to the stellar wind and, accordingly, a low loss rate of the intrinsic angular momentum of the stars at low metallicities give rise to a BH with a larger $\WI{\chi}{BH}$ (for a discussion, see \cite{Postnov2019}). As the metallicity decreases, $\WI{\chi}{BH}$ is shifted toward higher values: the maximum of this distribution is near zero for a solar metallicity and reaches $0.3$ for low $Z$ (see the right-hand panel of Fig.~\ref{Fig8}). This observation seems fairly important to us if we recall that the domain of admissible initial masses of the components at which the ejection of material will occur in the NS–BH system increases with $l{=}\WI{\chi}{BH}\cos{\WI{\theta}{BH}}$ (see Fig.~\ref{Fig4}).

The distribution of compact objects by masses refers us to the old, incompletely solved problem of calculating the core collapse SN \cite{MezzacappaFuller2005book}. Despite the significant progress of the last years in the field of three-dimensional collapse
calculations \cite{Burrows2020}, much unclear, in particular, the properties (and type) of the formed compact object, remains here. In view of the further probable progress in this field, in this paper we restricted ourselves to the following assumptions. The
mass of the forming BH was taken to be equal to the CO core mass before the progenitor collapse (as computed in the \texttt{BSE} code), given the 10\% of the gravitational mass defect, i.e., $M_\mathrm{BH}=0.9M_\mathrm{CO}$. The choice in favor of this distribution was made by taking into account the best agreement of the distribution of coalescing BHs by component masses and spins with the observational data from the LIGO-Virgo GW detectors \cite{Postnov2020}. To determine the masses of the forming NSs, we used the delayed collapse mechanism from \cite{Fryer2012} parameterized in the Appendix to the paper \cite{Giacobbo2018}.

\begin{figure*}[h]		
	\centering
	\includegraphics[width=\textwidth]{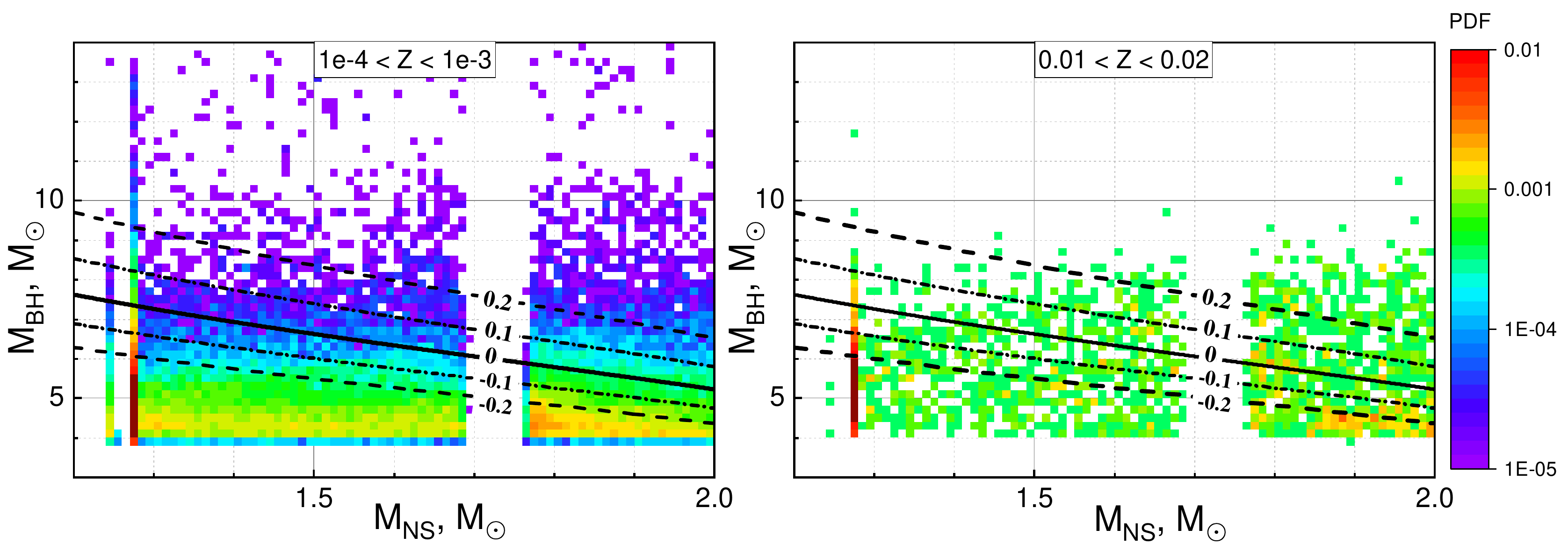}
	\caption{The distributions of coalescing NS–BH systems by component masses for the delayed NS formation mechanism from \cite{Fryer2012}. The lines indicate the boundaries between the NS plunge and stripping scenarios for various values of
	the specific BH spin projection ($l = \WI{\chi}{BH}cos{\WI{\theta}{BH}}$). Left-hand panel: low-metallicity ($10^{-4}<Z<10^{-3}$) stars. Right-hand panel: stars of nearly solar chemical composition ($0.01<Z<0.02$). NSs with masses of $1.25 \WI{M}{\odot}$ were formed as a result of the ECSN mechanism. }    
	\label{Fig9}
\end{figure*}

The distributions of coalescing NS–BH systems in component masses for various $Z$ are presented in Fig.~\ref{Fig9}. Note that using the model from \cite{Fryer2012} leads to a number of peculiarities:
\begin{itemize}
	\item near $\mNS{\approx} 1.7 M_\odot$ there is a small gap due to the peculiarity of the core mass growth near $\WI{M}{CO}{\approx} 3.5 M_\odot$ (see, e.g., \cite{Xing2023}), but the existence of such a gap is not confirmed by the observational data;
	\item there is also a distinct maximum near the lower NS mass boundary $\mNS{\approx} 1.28 M_\odot$, associated with the contribution of the stars from
	the evolutionary scenario with unstable mass transfer (CE1). In this case, the secondary component does not gain any additional mass in the primary mass transfer episode, and a large fraction of NSs are formed near the lower mass boundary.
\end{itemize}  

Therefore, as an alternative we also considered the model with a uniform mass distribution of NSs forming in the range $1.4 M_\odot \leq \mNS \leq 2.0 M_\odot$ (for observational justifications, see, e.g., \cite{Ozel2012}). It turned out that the peculiarities of the chosen NS formation mechanism do not affect significantly the final result of our calculations: most of the systems are in the domain of parameters leading to subsequent NS stripping, as can be qualitatively seen in Fig.~\ref{Fig9}.

\section{RESULTS}

In this section we will list the main results obtained during our joint analytical study of the boundaries between the NS plunge, tidal disruption, and stripping
scenarios as well as our population analysis of the evolution of massive close binary systems — NS–BH progenitors.

\begin{figure}[h]		
	\centering
	\includegraphics[scale=0.55]{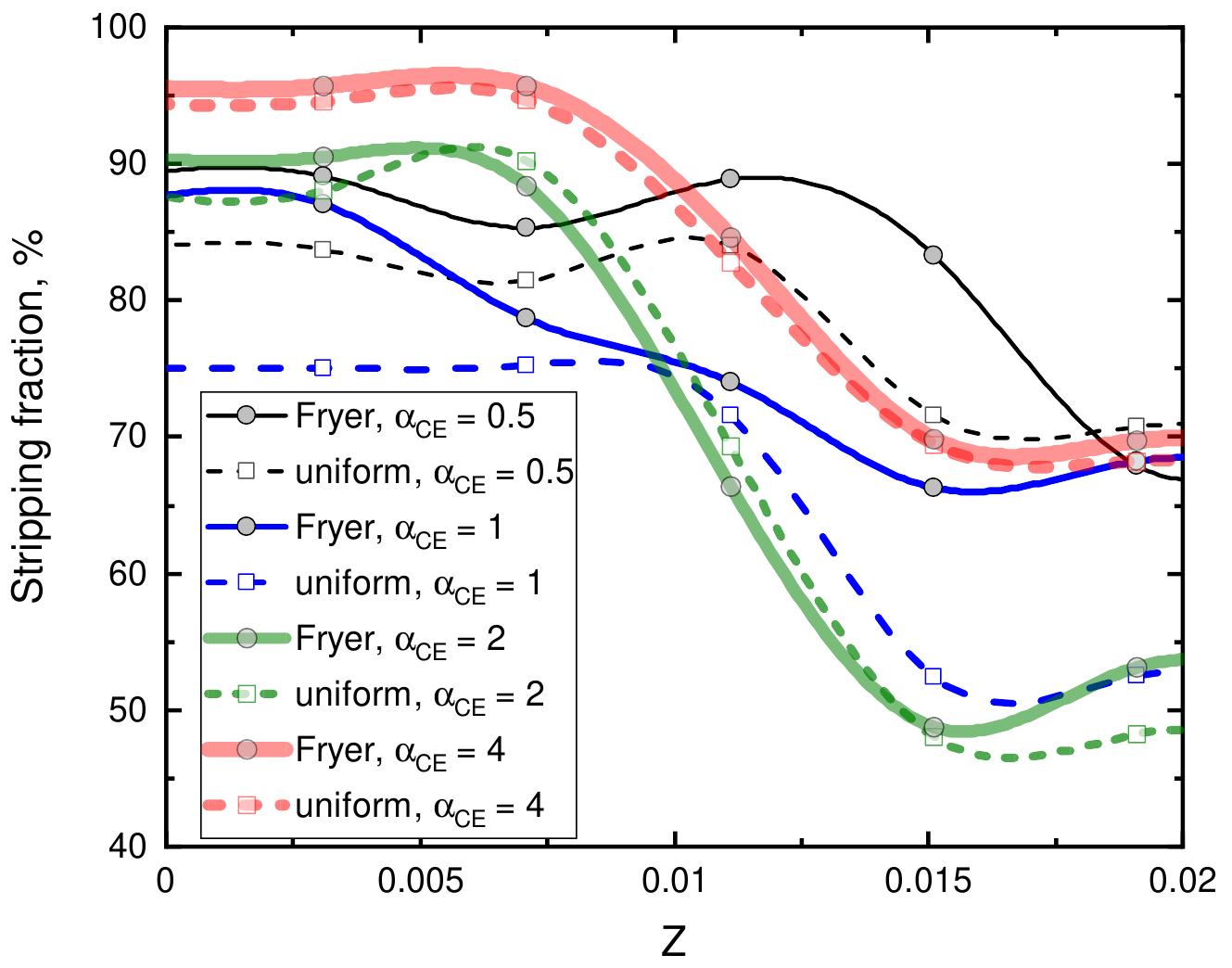} %
	\caption{Fraction of the stripping mechanism among all coalescing NS–BH systems as a function of the initial metallicity of the stellar population, depending on the model assumptions about the evolution of close binary systems. Different colors
	correspond to different values of the common-envelope efficiency parameter ($\WI{\alpha}{CE}$). The signs and line types on the graph
	specify the model of the NS mass distribution being used: the delayed mechanism from \cite{Fryer2012} (circles) and the uniform distribution (squares).}    
	\label{Fig10}
\end{figure}

\begin{itemize}
	
	\item Including the GR effects leads to a significant increase of the domain of initial NS and BH masses at which the stripping mechanism is realized, as can be clearly seen in Fig.~\ref{Fig2}. The relativistic mass transfer stability criterion obtained by us in the first and second PN approximations
	(Eqs.~(\ref{stab_PN1}) and (\ref{Appendix_stab_PN2}), respectively) narrows greatly the domain of initial parameters of the system at which the NS, having
	avoided the plunge into the BH, will be subjected to rapid tidal disruption. Instead long BH stripping awaits the NS, which disagrees, for example, with the purely Newtonian estimate from Appendix~C.1 of the review \cite{Kyutoku2021}, that rejects any possibility of such an outcome.
	
	\item As the BH spin projection onto the system’s orbital angular momentum axis grows, the domain of initial component masses at which the ejection of material is possible increases significantly (see Fig.~\ref{Fig4}), consistent with similar
	analytical estimates (Eq.~(9) from \cite{Kyutoku2021}) and hydrodynamic simulations \cite{Kyutoku2015}. This is related to the so-called spin–orbit coupling that arises in the 1.5PN approximation. Note that the boundary between the NS stripping and tidal disruption scenarios is virtually insensitive to this effect.
	
	\item Other things being equal, the uncertainty in choosing the EoS affects weakly the mass boundary between the NS plunge and stripping scenarios (see Fig.~\ref{Fig5}) and has virtually no effect on the mass transfer stability, as discussed in detail in \cite{KramarevYudin2023str} devoted to binary NSs. 
	
	\item To discuss the results of our population calculations, let us turn to Fig.~\ref{Fig10}, where the fraction of stripping NS–BH systems is presented as
	a function of the metallicity of stars from the original population. The fraction of the tidal disruption mechanism in all cases is $\lesssim 0.1\%$, and, therefore, in the remaining systems the NS entirely plunges below the BH horizon at the end. It can be clearly seen from this figure that, other things being equal, an increase in $Z$ leads to a decrease in the fraction of the stripping mechanism and, consequently, to an increase in the fraction of the NS plunge scenario.
	There are two reasons for this. First, the stellar wind intensity (see, e.g., \cite{Kudritzki2000}) and, along with it, the loss rate of the intrinsic angular momentum by the BH progenitor star increase with metallicity. Hence, the final BH spin, $\WI{\chi}{BH}$, will be lower (see Fig.~\ref{Fig8}). Second, as the metallicity increases, by the time of the first SN explosion the fraction of wide binary systems that experience a significant rotation of the orbital plane after
	this explosion grows (Fig.~\ref{Fig7}). Consequently, for such systems the cosine of the BH spin inclination angle to the orbital angular momentum
	vector, $\cos{\WI{\theta}{BH}}$, on average, will be smaller. A combination of these factors leads to a decrease of the projection of the intrinsic
	BH angular momentum, $l{=}\WI{\chi}{BH} \cos{\WI{\theta}{BH}}$, and, along with it, the domain of initial NS and BH masses at which the NS stripping mechanism is realized.
	
	\item Although small variations in the parameter $\WI{\alpha}{CE}$ are capable of changing the total number of coalescing NS–BH systems by an order of magnitude, the fraction of the stripping mechanism is weakly sensitive to its exact value: the changes are only a few tens of percent, as can be seen from Fig.~\ref{Fig10}.
	
	\item To investigate the influence of the NS mass distribution after the SN explosion, we considered two fairly different model assumptions: the delayed mechanism from \cite{Fryer2012} and the uniform distribution in the range of
	masses $1.4 M_\odot \leq \mNS \leq 2.0 M_\odot$ (see Fig.~\ref{Fig10}). The BH mass after the collapse in both cases was taken to be equal to the CO core mass of
	the progenitor star (corrected for the gravitational mass defect). Our calculations clearly show that, other things being equal, the model of the NS mass distribution affects weakly the fraction of the stripping mechanism.
	
\end{itemize}  

\section{CONCLUSIONS AND FURTHER PROSPECTS}

The NS stripping mechanism was initially developed in the context of coalescing binary NSs with a significant asymmetry of the component masses \cite{Blinnikov2021}. This was one of the reasons that prompted us to also consider this mechanism for NS–BH systems, where the component masses, as a rule, differ greatly from each other (see \cite{Broekgaarden2021} and references therein). However, despite the significant mass asymmetry in NS–BH systems, there were serious doubts in the existence of stable mass transfer related to the action of GR effects even at the analytical level of consideration. Nevertheless, our analysis showed that, given the relativistic
effects, the stripping mechanism is not just possible, but can be the dominant scenario for the final evolution of such systems.

A so optimistic result motivates us to continue the started study in future. Within the analytical approach being developed, we plan to take into account the tidal effects that can affect significantly the mass transfer stability. Thereafter, we are going
to calculate the potentially observed number of systems with NS stripping and subsequent explosion by population synthesis methods. To calculate the necessary characteristics of the EM signal (including the kilonova) from a minimum-mass exploding NS, in future we plan to use the \texttt{STELLA} code \cite{Blinnikov2006} by first modifying it to take into account the transuranium elements.\footnote{For preliminary estimates one can use approximation expressions, for example, Eq.~(12) for the luminosity from \cite{Kawaguchi2016}.} Note that the limiting sensitivity of the LIGO–Virgo facilities allows one to detect GW alerts from coalescing compact objects
at distances $d\simeq200 \, \text{Mpc}$, corresponding to galaxies with nearly solar metallicities (see, e.g., Fig.~6 from \cite{Gallazzi2005}). In such stellar populations
the stripping mechanism is realized in 50–70\% of the cases, depending on the model assumptions about the evolution of binary systems being used (the right region in Fig.~\ref{Fig10}). Given also the relatively large mass of the ejection of neutron-rich material in such events, their further prospects of detection seem to us fairly
promising.

At the same time, one cannot but mention that up-to-date hydrodynamic simulations of NS–BH mergers dismiss the possibility of NS stripping (see \cite{Kyutoku2021} and references therein). Note that this may be related to the improper selection of initial
conditions in hydrodynamic simulations, as, for example, was shown in \cite{Dan2011} for binary white dwarfs and in \cite{Blinnikov2022} for binary NSs. In this sense the analytical study presented in this paper is also a good motivation for more careful
hydrodynamic simulations of mergers/stripping in NS–BH systems. We plan to perform such simulations using the open hydrodynamic SPH-code \texttt{PHANTOM} \cite{Price2018} in the near future. 

After the submission of this paper, it became known that the GW signal was recorded from a coalescing NS–BH system with a possible anomalously low BH mass \cite{LVK2024}. In light of this event, in further population calculations it is necessary to additionally study the influence of the lower boundary of the BH mass distribution on the fraction of each scenario.

\section*{\textit{APPENDIX.} THE MASS TRANSFER STABILITY CRITERION IN THE SECOND POST-NEWTONIAN APPROXIMATION}

In this Appendix we will present the mass transfer stability criterion in the 2PN approximation by taking into account the 1.5PN-order spin–orbit coupling. For convenience, let us first introduce the notation for the coefficients of the high-order PN corrections (proportional to $z^{1.5}$ and $z^2$) in Eq.~(\ref{J_orb}) for the orbital angular momentum:
\begin{equation}
	\WI{\alpha}{SO}{=}\frac{3}{2\sqrt{8}}\left(2{+}\frac{3}{2}q\right)\frac{\nu l}{q}, \; \WI{\alpha}{2}{=}\frac{42{-}43\nu}{64}. \label{Appendix_1}  
\end{equation}
We will also need their dimensionless derivatives with respect to the NS mass:
\begin{eqnarray}
	\WI{\alpha}{SO}^{\mathrm{M}}\equiv \frac{d\WI{\alpha}{SO}}{d\mNS}\mNS= \frac{9}{4\sqrt{8}}(1{+}q)\nu l - \frac{3}{2\sqrt{8}} \, \frac{4q{+}3 q^2}{2(1{+}q)} \, \mathfrak{j} \, \sqrt{\frac{2}{z}} \left(\cos{\WI{\theta}{BH}}{+}\sin^2{\WI{\theta}{BH}}\right), \label{Appendix_2}  
\end{eqnarray}
\begin{equation}
	\WI{\alpha}{2}^{\mathrm{M}}\equiv \frac{d\WI{\alpha}{2}}{d\mNS}\mNS= -\frac{43}{64} (1{-}q) \nu.\label{Appendix_3}  
\end{equation}
When deriving Eq.~(\ref{Appendix_2}), we took into account the change in the BH spin projection, $l{=}\WI{\chi}{BH} \cos{\WI{\theta}{BH}}$, using the accretion spin-up formula~(\ref{J_BH}) for $\Delta \WI{\chi}{BH} {\sim}\Delta\WI{J}{BH}$ and the expression $\Delta\cos{\WI{\theta}{BH}}{=}\sin^2{\WI{\theta}{BH}} \frac{\Delta \WI{J}{BH}}{\WI{J}{BH}}$. With the notation introduced above the mass transfer stability criterion in the 2PN approximation can be written as
\begin{eqnarray}
	\frac{d \ln \rNS}{d \ln \mNS} \geq (1{+}q)\left[\frac{d \ln \WI{f}{Egg}}{d \ln q}+\frac{zq}{1{+}z\WI{f}{z}}\frac{d \WI{f}{z}}{d q}\right]   \nonumber \\ -\frac{2\left[(1{-}q)(1+z-\WI{\alpha}{SO}z^{1.5}+\WI{\alpha}{2}z^2)\right]}{\left[1-z+2\WI{\alpha}{SO}z^{1.5}-3\WI{\alpha}{2}z^2\right](1{+}z\WI{f}{z})} 
	+  \frac{2\left[\mathfrak{j}(1{+}q)+\WI{\alpha}{SO}^{\mathrm{M}}z^{1.5}-\WI{\alpha}{2}^{\mathrm{M}}z^2\right]}{\left[1-z+2\WI{\alpha}{SO}z^{1.5}-3\WI{\alpha}{2}z^2\right](1{+}z\WI{f}{z})}. 
	\label{Appendix_stab_PN2}  
\end{eqnarray}
In the 1PN approximation this expression transforms into Eq.~(\ref{stab_PN1}).

\section*{ACKNOWLEDGEMENTS}

The work of N.I.~Kramarev was supported by the ''BASIS'' Foundation for the Development of Theoretical Physics and Mathematics (project no.~22-2-10-11-1). The work of A.G.~Kuranov was performed using the equipment of the Data Storage and Processing System of the Sternberg Astronomical Institute of the Moscow State University purchased through the funds of the Program for the Development of the Moscow State University.
A.V.~Yudin thanks the Russian Science Foundation (grant no.~22-12-00103) for its support.

\bibliographystyle{unsrt}  
\bibliography{references}

\begin{thebibliography}{10}

\bibitem{LattimerSchramm1974}
J.~M. {Lattimer} and D.~N. {Schramm}.
\newblock {Black-Hole-Neutron-Star Collisions}.
\newblock {\em \apjl}, 192:L145, September 1974.

\bibitem{ClarkEardley1977}
J.~P.~A. {Clark} and D.~M. {Eardley}.
\newblock {Evolution of close neutron star binaries.}
\newblock {\em \apj}, 215:311--322, July 1977.

\bibitem{Blinnikov1984}
S.~I. {Blinnikov}, I.~D. {Novikov}, T.~V. {Perevodchikova}, and A.~G.
  {Polnarev}.
\newblock {Exploding Neutron Stars in Close Binaries}.
\newblock {\em Soviet Astronomy Letters}, 10:177--179, April 1984.

\bibitem{Eichler1989}
D.~{Eichler}, M.~{Livio}, T.~{Piran}, and D.~N. {Schramm}.
\newblock {Nucleosynthesis, neutrino bursts and {\ensuremath{\gamma}}-rays from
  coalescing neutron stars}.
\newblock {\em Nature}, 340(6229):126--128, July 1989.

\bibitem{Li1998}
L.-X. {Li} and B.~{Paczy{\'n}ski}.
\newblock {Transient Events from Neutron Star Mergers}.
\newblock {\em \apjl}, 507(1):L59--L62, November 1998.

\bibitem{Metzger2010}
B.~D. {Metzger}, G.~{Mart{\'\i}nez-Pinedo}, S.~{Darbha}, and E.~{Quataert, et
  al.}
\newblock {Electromagnetic counterparts of compact object mergers powered by
  the radioactive decay of r-process nuclei}.
\newblock {\em \mnras}, 406(4):2650--2662, August 2010.

\bibitem{Abbott2017}
B.~P. {Abbott}, R.~{Abbott}, T.~D. {Abbott}, and F.~{Acernese, et al.}
\newblock {Gravitational Waves and Gamma-Rays from a Binary Neutron Star
  Merger: GW170817 and GRB 170817A}.
\newblock {\em \apjl}, 848(2):L13, October 2017.

\bibitem{Villar2017}
V.~A. {Villar}, J.~{Guillochon}, E.~{Berger}, and B.~D. {Metzger, et al.}
\newblock {The Combined Ultraviolet, Optical, and Near-infrared Light Curves of
  the Kilonova Associated with the Binary Neutron Star Merger GW170817: Unified
  Data Set, Analytic Models, and Physical Implications}.
\newblock {\em \apjl}, 851(1):L21, December 2017.

\bibitem{Bhattacharya2019}
M.~{Bhattacharya}, P.~{Kumar}, and G.~{Smoot}.
\newblock {Mergers of black hole-neutron star binaries and rates of associated
  electromagnetic counterparts}.
\newblock {\em \mnras}, 486(4):5289--5309, July 2019.

\bibitem{Ekanger2023}
N.~{Ekanger}, M.~{Bhattacharya}, and S.~{Horiuchi}.
\newblock {Nucleosynthesis in outflows of compact objects and detection
  prospects of associated kilonovae}.
\newblock {\em \mnras}, 525(2):2040--2052, October 2023.

\bibitem{Drozda2022}
P.~{Drozda}, K.~{Belczynski}, R.~{O'Shaughnessy}, and T.~{Bulik, et al.}
\newblock {Black hole-neutron star mergers: The first mass gap and kilonovae}.
\newblock {\em \aap}, 667:A126, November 2022.

\bibitem{Postnov2020}
K.~A. {Postnov}, A.~G. {Kuranov}, and I.~V. {Simkin}.
\newblock {Possible Electromagnetic Phenomena during the Coalescence of Neutron
  Star-Black Hole Binary Systems}.
\newblock {\em Astronomy Letters}, 45(11):728--739, February 2020.

\bibitem{Kyutoku2021}
K.~{Kyutoku}, M.~{Shibata}, and K.~{Taniguchi}.
\newblock {Coalescence of black hole-neutron star binaries}.
\newblock {\em Living Reviews in Relativity}, 24(1):5, December 2021.

\bibitem{Blinnikov2021}
S.~I. {Blinnikov}, D.~K. {Nadyozhin}, N.~I. {Kramarev}, and A.~V. {Yudin}.
\newblock {Neutron Star Mergers and Gamma-Ray Bursts: Stripping Model}.
\newblock {\em Astronomy Reports}, 65(5):385--391, May 2021.

\bibitem{Blinnikov2022}
S.~{Blinnikov}, A.~{Yudin}, N.~{Kramarev}, and M.~{Potashov}.
\newblock {Stripping Model for Short Gamma-Ray Bursts in Neutron Star Mergers}.
\newblock {\em Particles}, 5(2):198--209, June 2022.

\bibitem{KramarevYudin2023str}
N.~{Kramarev} and A.~{Yudin}.
\newblock {Accretion spin-up of the massive component in the neutron star
  stripping model for short gamma-ray bursts}.
\newblock {\em \mnras}, 525(3):3306--3315, November 2023.

\bibitem{Blinnikov1990}
S.~I. {Blinnikov}, V.~S. {Imshennik}, D.~K. {Nadezhin}, and I.~D. {Novikov, et
  al.}
\newblock {Explosion of a Low-Mass Neutron Star}.
\newblock {\em Soviet Astronomy}, 34:595, December 1990.

\bibitem{Panov2020}
I.~V. {Panov} and A.~V. {Yudin}.
\newblock {Production of Heavy Elements during the Explosion of a Low-Mass
  Neutron Star in a Close Binary}.
\newblock {\em Astronomy Letters}, 46(8):518--527, August 2020.

\bibitem{Yudin2022}
A.~V. {Yudin}.
\newblock {Explosion of a Minimum-Mass Neutron Star within Relativistic
  Hydrodynamics}.
\newblock {\em Astronomy Letters}, 48(6):311--320, June 2022.

\bibitem{Yip2022}
C.-M. {Yip}, M.-C. {Chu}, S.-C. {Leung}, and L.-M. {Lin}.
\newblock {R-process Nucleosynthesis of Subminimal Neutron Star Explosions}.
\newblock {\em \apj}, 956(2):115, October 2023.

\bibitem{Kawaguchi2016}
K.~{Kawaguchi}, K.~{Kyutoku}, M.~{Shibata}, and M.~{Tanaka}.
\newblock {Models of Kilonova/Macronova Emission from Black Hole-Neutron Star
  Mergers}.
\newblock {\em \apj}, 825(1):52, July 2016.

\bibitem{Lovelace2013}
G.~{Lovelace}, M.~D. {Duez}, F.~{Foucart}, and L.~E. {Kidder, et al.}
\newblock {Massive disc formation in the tidal disruption of a neutron star by
  a nearly extremal black hole}.
\newblock {\em Classical and Quantum Gravity}, 30(13):135004, July 2013.

\bibitem{Kiuchi2015}
K.~{Kiuchi}, Y.~{Sekiguchi}, K.~{Kyutoku}, and M.~{Shibata, et al.}
\newblock {High resolution magnetohydrodynamic simulation of black hole-neutron
  star merger: Mass ejection and short gamma ray bursts}.
\newblock {\em \prd}, 92(6):064034, September 2015.

\bibitem{Hayashi2021}
K.~{Hayashi}, K.~{Kawaguchi}, K.~{Kiuchi}, and K.~{Kyutoku, et al.}
\newblock {Properties of the remnant disk and the dynamical ejecta produced in
  low-mass black hole-neutron star mergers}.
\newblock {\em \prd}, 103(4):043007, February 2021.

\bibitem{Hayashi2022}
K.~{Hayashi}, S.~{Fujibayashi}, K.~{Kiuchi}, and K.~{Kyutoku, et al.}
\newblock {General-relativistic neutrino-radiation magnetohydrodynamic
  simulation of seconds-long black hole-neutron star mergers}.
\newblock {\em \prd}, 106(2):023008, July 2022.

\bibitem{Xing2023}
Z.~{Xing}, S.~S. {Bavera}, T.~{Fragos}, and M.~U. {Kruckow, et al.}
\newblock {From ZAMS to Merger: Detailed Binary Evolution Models of Coalescing
  Neutron Star-Black Hole Systems at Solar Metallicity}.
\newblock {\em arXiv e-prints}, page arXiv:2309.09600, September 2023.

\bibitem{Zhu2022}
J.-P. {Zhu}, S.~{Wu}, Y.~{Qin}, and B.~{Zhang, et al.}
\newblock {Population Properties of Gravitational-wave Neutron Star-Black Hole
  Mergers}.
\newblock {\em \apj}, 928(2):167, April 2022.

\bibitem{Kowalska2011}
I.~{Kowalska}, T.~{Bulik}, K.~{Belczynski}, and M.~{Dominik, et al.}
\newblock {The eccentricity distribution of compact binaries}.
\newblock {\em \aap}, 527:A70, March 2011.

\bibitem{Davies2005}
M.~B. {Davies}, A.~J. {Levan}, and A.~R. {King}.
\newblock {The ultimate outcome of black hole-neutron star mergers}.
\newblock {\em \mnras}, 356(1):54--58, January 2005.

\bibitem{Rezzolla2016}
L.~{Rezzolla}.
\newblock {\em An Introduction to Astrophysical Black Holes and Their Dynamical
  Production}, pages 1--44.
\newblock Springer International Publishing, Cham, 2016.

\bibitem{Bardeen1972}
J.~M. {Bardeen}, W.~H. {Press}, and S.~A. {Teukolsky}.
\newblock {Rotating Black Holes: Locally Nonrotating Frames, Energy Extraction,
  and Scalar Synchrotron Radiation}.
\newblock {\em \apj}, 178:347--370, December 1972.

\bibitem{Gompertz2022}
B.~P. {Gompertz}, M.~{Nicholl}, P.~{Schmidt}, and G.~{Pratten, et al.}
\newblock {Constraints on compact binary merger evolution from spin-orbit
  misalignment in gravitational-wave observations}.
\newblock {\em \mnras}, 511(1):1454--1461, March 2022.

\bibitem{Eggleton1983}
P.~P. {Eggleton}.
\newblock {Aproximations to the radii of Roche lobes.}
\newblock {\em \apj}, 268:368--369, May 1983.

\bibitem{Ratkovic2005}
S.~{Ratkovic}, M.~{Prakash}, and J.~M. {Lattimer}.
\newblock {Roche Lobes in the Second Post-Newtonian Approximation}.
\newblock {\em arXiv e-prints}, pages astro--ph/0512133, December 2005.

\bibitem{Jaranowsk1992J}
P.~{Jaranowski} and A.~{Krolak}.
\newblock {Detectability of the Gravitational Wave Signal from a Close Neutron
  Star Binary with Mass Transfer}.
\newblock {\em \apj}, 394:586, August 1992.

\bibitem{PortegiesZwart1998}
S.~F. {Portegies Zwart}.
\newblock {Gamma-Ray Binaries: Stable Mass Transfer from a Neutron Star to a
  Black Hole}.
\newblock {\em \apjl}, 503:L53, August 1998.

\bibitem{Paczynski1969}
B.~{Paczy{\'n}ski}, J.~{Zi{\'o}lkowski}, and A.~{Zytkow}.
\newblock {On the Time-Scale of the Mass Transfer in Close Binaries}.
\newblock In Margherita {Hack}, editor, {\em Mass Loss from Stars}, volume~13
  of {\em Astrophysics and Space Science Library}, page 237, January 1969.

\bibitem{Paczynski1967}
B.~{Paczy{\'n}ski}.
\newblock {Gravitational Waves and the Evolution of Close Binaries}.
\newblock {\em Acta Astronomica}, 17:287, January 1967.

\bibitem{SchaferWex1993}
G.~{Sch{\"a}fer} and N.~{Wex}.
\newblock {Second post-Newtonian motion of compact binaries}.
\newblock {\em Physics Letters A}, 174(3):196--205, March 1993.

\bibitem{Wex1995}
N.~{Wex}.
\newblock {The second post-Newtonian motion of compact binary-star systems with
  spin}.
\newblock {\em Classical and Quantum Gravity}, 12(4):983--1005, April 1995.

\bibitem{KramarevYudin2023acc}
N.~{Kramarev} and A.~{Yudin}.
\newblock {Dynamics of direct impact accretion in degenerate binary systems}.
\newblock {\em \mnras}, 522(1):626--634, June 2023.

\bibitem{LubowShu1975}
S.~H. {Lubow} and F.~H. {Shu}.
\newblock {Gas dynamics of semidetached binaries.}
\newblock {\em \apj}, 198:383--405, June 1975.

\bibitem{Kyutoku2015}
K.~{Kyutoku}, K.~{Ioka}, H.~{Okawa}, and M.~{Shibata, et al.}
\newblock {Dynamical mass ejection from black hole-neutron star binaries}.
\newblock {\em \prd}, 92(4):044028, August 2015.

\bibitem{Pearson2018}
J.~M. {Pearson}, N.~{Chamel}, A.~Y. {Potekhin}, and A.~F. {Fantina, et al.}
\newblock {Unified equations of state for cold non-accreting neutron stars with
  Brussels-Montreal functionals - I. Role of symmetry energy}.
\newblock {\em \mnras}, 481(3):2994--3026, December 2018.

\bibitem{Raaijmakers2020}
G.~{Raaijmakers}, S.~K. {Greif}, T.~E. {Riley}, and T.~{Hinderer, et al.}
\newblock {Constraining the Dense Matter Equation of State with Joint Analysis
  of NICER and LIGO/Virgo Measurements}.
\newblock {\em \apjl}, 893(1):L21, April 2020.

\bibitem{Greif2020}
S.~K. {Greif}, K.~{Hebeler}, J.~M. {Lattimer}, and C.~J. {Pethick, et al.}
\newblock {Equation of State Constraints from Nuclear Physics, Neutron Star
  Masses, and Future Moment of Inertia Measurements}.
\newblock {\em \apj}, 901(2):155, October 2020.

\bibitem{Taniguchi2008}
K.~{Taniguchi}, T.~W. {Baumgarte}, J.~A. {Faber}, and S.~L. {Shapiro}.
\newblock {Relativistic black hole-neutron star binaries in quasiequilibrium:
  Effects of the black hole excision boundary condition}.
\newblock {\em \prd}, 77(4):044003, February 2008.

\bibitem{Postnov2014}
K.~A. {Postnov} and L.~R. {Yungelson}.
\newblock {The Evolution of Compact Binary Star Systems}.
\newblock {\em Living Reviews in Relativity}, 17(1):3, May 2014.

\bibitem{Hurley2000}
J.~R. {Hurley}, O.~R. {Pols}, and C.~A. {Tout}.
\newblock {Comprehensive analytic formulae for stellar evolution as a function
  of mass and metallicity}.
\newblock {\em \mnras}, 315:543--569, July 2000.

\bibitem{Hurley2002}
J.~R. {Hurley}, C.~A. {Tout}, and O.~R. {Pols}.
\newblock {Evolution of binary stars and the effect of tides on binary
  populations}.
\newblock {\em \mnras}, 329:897--928, February 2002.

\bibitem{Postnov2019}
K.~A. {Postnov} and A.~G. {Kuranov}.
\newblock {Black hole spins in coalescing binary black holes}.
\newblock {\em \mnras}, 483(3):3288--3306, March 2019.

\bibitem{Broekgaarden2021}
F.~S. {Broekgaarden}, E.~{Berger}, C.~J. {Neijssel}, and A.~{Vigna-G{\'o}mez,
  et al.}
\newblock {Impact of massive binary star and cosmic evolution on gravitational
  wave observations I: black hole-neutron star mergers}.
\newblock {\em \mnras}, 508(4):5028--5063, December 2021.

\bibitem{Webbink1984}
R.~F. {Webbink}.
\newblock {Double white dwarfs as progenitors of R Coronae Borealis stars and
  Type I supernovae}.
\newblock {\em \apj}, 277:355--360, February 1984.

\bibitem{Iben1984}
I.~{Iben, Jr.,} and A.~V. {Tutukov}.
\newblock {Supernovae of type I as end products of the evolution of binaries
  with components of moderate initial mass (M not greater than about 9 solar
  masses)}.
\newblock {\em \apjs}, 54:335--372, February 1984.

\bibitem{Fryer2012}
C.~L. {Fryer}, K.~{Belczynski}, G.~{Wiktorowicz}, and M.~{Dominik, et al.}
\newblock {Compact Remnant Mass Function: Dependence on the Explosion Mechanism
  and Metallicity}.
\newblock {\em \apj}, 749(1):91, April 2012.

\bibitem{Kudritzki2000}
R.-P. {Kudritzki} and J.~{Puls}.
\newblock {Winds from Hot Stars}.
\newblock {\em \araa}, 38:613--666, January 2000.

\bibitem{Kippenhahn1967}
R.~{Kippenhahn} and A.~{Weigert}.
\newblock {Entwicklung in engen Doppelsternsystemen I. Massenaustausch vor und
  nach Beendigung des zentralen Wasserstoff-Brennens}.
\newblock {\em Zeitschrift für Astrophysik}, 65:251, January 1967.

\bibitem{deKool1990}
M.~{de Kool}.
\newblock {Common Envelope Evolution and Double Cores of Planetary Nebulae}.
\newblock {\em \apj}, 358:189, July 1990.

\bibitem{Loveridge2011}
A.~J. {Loveridge}, M.~V. {van der Sluys}, and V.~{Kalogera}.
\newblock {Analytical Expressions for the Envelope Binding Energy of Giants as
  a Function of Basic Stellar Parameters}.
\newblock {\em \apj}, 743(1):49, Dec 2011.

\bibitem{Hobbs2005}
G.~{Hobbs}, D.~R. {Lorimer}, A.~G. {Lyne}, and M.~{Kramer}.
\newblock {A statistical study of 233 pulsar proper motions}.
\newblock {\em \mnras}, 360(3):974--992, July 2005.

\bibitem{Miyaji1980}
S.~{Miyaji}, K.~{Nomoto}, K.~{Yokoi}, and D.~{Sugimoto}.
\newblock {Supernova Triggered by Electron Captures}.
\newblock {\em \pasj}, 32:303, 1980.

\bibitem{Siess2018}
L.~{Siess} and U.~{Lebreuilly}.
\newblock {Case A and B evolution towards electron capture supernova}.
\newblock {\em \aap}, 614:A99, June 2018.

\bibitem{MezzacappaFuller2005book}
A.~Mezzacappa and G.~M. Fuller.
\newblock {\em Open Issues in Core Collapse Supernova Theory}.
\newblock WORLD SCIENTIFIC, 2005.

\bibitem{Burrows2020}
A.~{Burrows}, D.~{Radice}, D.~{Vartanyan}, and H.~{Nagakura, et al.}
\newblock {The overarching framework of core-collapse supernova explosions as
  revealed by 3D FORNAX simulations}.
\newblock {\em \mnras}, 491(2):2715--2735, January 2020.

\bibitem{Giacobbo2018}
N.~{Giacobbo}, M.~{Mapelli}, and M.~{Spera}.
\newblock {Merging black hole binaries: the effects of progenitor's
  metallicity, mass-loss rate and Eddington factor}.
\newblock {\em \mnras}, 474:2959--2974, March 2018.

\bibitem{Ozel2012}
F.~{{\"O}zel}, D.~{Psaltis}, R.~{Narayan}, and A.~{Santos Villarreal}.
\newblock {On the Mass Distribution and Birth Masses of Neutron Stars}.
\newblock {\em \apj}, 757(1):55, September 2012.

\bibitem{Blinnikov2006}
S.~I. {Blinnikov}, F.~K. {R{\"o}pke}, E.~I. {Sorokina}, and M.~{Gieseler, et
  al.}
\newblock {Theoretical light curves for deflagration models of type Ia
  supernova}.
\newblock {\em \aap}, 453(1):229--240, July 2006.

\bibitem{Gallazzi2005}
A.~{Gallazzi}, S.~{Charlot}, J.~{Brinchmann}, and S.~D.~M. {White, et al.}
\newblock {The ages and metallicities of galaxies in the local universe}.
\newblock {\em \mnras}, 362(1):41--58, September 2005.

\bibitem{Dan2011}
M.~{Dan}, S.~{Rosswog}, J.~{Guillochon}, and E.~{Ramirez-Ruiz}.
\newblock {Prelude to A Double Degenerate Merger: The Onset of Mass Transfer
  and Its Impact on Gravitational Waves and Surface Detonations}.
\newblock {\em \apj}, 737(2):89, August 2011.

\bibitem{Price2018}
D.~J. {Price}, J.~{Wurster}, T.~S. {Tricco}, and C.~{Nixon, et al.}
\newblock {Phantom: A Smoothed Particle Hydrodynamics and Magnetohydrodynamics
  Code for Astrophysics}.
\newblock {\em Publications of the Astronomical Society of Australia}, 35:e031,
  September 2018.

\bibitem{LVK2024}
A.~G. {Abac}, R.~{Abbott}, I.~{Abouelfettouh}, and F.~{Acernese, et al.}
\newblock {Observation of Gravitational Waves from the Coalescence of a
  2.5{\textendash}4.5 M $_{{\ensuremath{\odot}}}$ Compact Object and a Neutron
  Star}.
\newblock {\em \apjl}, 970(2):L34, August 2024.

\end{thebibliography}

\end{document}